\documentclass{article}
\usepackage[utf8]{inputenc}
\usepackage{graphicx, amsmath}

\usepackage[left=2cm, right=2cm, top=2cm]{geometry}
\usepackage{graphicx}
\usepackage{tabularx}
\usepackage{tabu}
\usepackage{longtable}
\usepackage{hyperref}
\usepackage{siunitx}
\DeclareMathOperator\erf{erf}
\usepackage{authblk}
\usepackage{subcaption}
\usepackage[export]{adjustbox} 
\usepackage[backend=biber, maxnames=1, minnames=1, style = numeric, sorting = none]{biblatex}
\title{\textbf{Optimizing for a Near Single-Mode Type-0 Optical Parametric Amplifier in Nanophotonics}}
\author[1]{Shivam Mundhra}
\author[2]{Elina Sendonaris}
\author[3]{Robert M. Gray}
\author[3]{James Williams}
\author[2,3]{Alireza Marandi}
\affil[1]{\small The Department of Physics, University of Chicago, Chicago, IL 60637, USA}
\affil[2]{\small Department of Applied Physics, California Institute of Technology, Pasadena, CA 91125, USA}
\affil[3]{\small Department of Electrical Engineering, California Institute of Technology, Pasadena, CA 91125, USA}

\begin{document}

\date{}

\maketitle

\begin{abstract}

Thin-film lithium niobate (TFLN) has recently emerged as a promising platform for integrated nonlinear photonics, enabling the use of optical parametric amplifiers (OPAs) for applications in quantum information processing, precision metrology, and ultrafast optical signal processing. However, OPA waveguide designs have not yet achieved the phase-matching conditions for type-0 operation in a single spectro-temporal mode, limiting their use. We optimize the waveguide dimensions, poling pattern, pump wavelength, and pump pulse duration for high spectral purity, a metric for single-mode fidelity. We numerically demonstrate a nanophotonic OPA with a spectral purity of 0.982 in a TFLN waveguide. Through semi-classical simulations, we further demonstrate that in the optical parametric regime, where vacuum fluctuations at the input of the OPA can saturate the gain and deplete the pump, the macroscopic output of such a single-mode OPA can be utilized for an ultra-fast quantum random number generator. These results demonstrate a promising direction for integrated OPAs in a wide range of ultrafast quantum nanophotonics applications.

\end{abstract}

\section{Introduction}

Integrated optical parametric amplifiers (OPAs) have recently garnered attention due to their demonstrated ability to generate squeezed vacuum states \cite{Taguchi2022, nehra2022} and a proposal for their application in quantum nondemolition measurements \cite{ryotatsu2023}. However, the energy and momentum conservation laws governing the OPA process induce spectral correlations between the signal and idler outputs \cite{Kuo16}, which can cause multimodal operation in the ultrashort-pulse regime \cite{multimodal}. This reduces the purity of the generated quantum states, limiting their use for quantum information applications. Dispersion and poling engineering can be employed to minimize the number of spectro-temporal modes present in the output states \cite{Xin2022}, but single-mode OPAs are still challenging to design and realize with type-0 phase matching, where the pump and signal share the same polarization.

We are particularly interested in using the single-mode output of an OPA as a source for quantum random number generators (QRNGs). The key mechanism for this process is the depletion of the pump by optical parametric generation (OPG), in which vacuum fluctuations are amplified to high brightness \cite{jankowski2022,ledezma2022}. In the single-mode regime, the macroscopic OPG output is expected to exhibit a binary phase behavior, with an approximately $50\%$ likelihood of adopting either a 0 phase or a $\pi$ phase, similar to the operation of degenerate optical parametric oscillators (OPOs) \cite{marandi2012}. By mapping these phases to binary states — mapping the 0 phase to logical 0 and the $\pi$ phase to logical 1 — the single-mode output can be leveraged as a QRNG. QRNGs can function as robust stochastic sources for various advanced computing applications \cite{ghahramani2015, pbits, misra2023}. They can also be useful for Ising machines and quantum Monte Carlo solvers \cite{mohseni2022, chowdhury2023}, and for strengthening the security of cryptographic protocols \cite{cryptography2011, bouda2012,li2015}. Previous experiments have explored the generation of probabilistic bits using the vacuum field of OPOs \cite{marandi2012, roques2023}. We aim to extend this generation of random bits to OPA waveguides, which can be realized in a much simpler configuration than OPOs and can operate in the ultrafast regime where the separation of the random bits can be on the order of femtoseconds.


In this study, we optimize a type-0 thin-film lithium niobate (TFLN) OPA and achieve a simulated output spectral purity of $0.982$, a measure of the dominance of the primary spectro-temporal mode. We apply our simulated waveguide to the realization of a QRNG. Using semi-classical nonlinear time-domain simulations \cite{ledezma2022} to simulate the OPA in the ultrafast pulsed regime, we verify that the phases of the output pulses can be resolved and mapped to a binary variable. Specifically, we fit our processed output fields to a bimodal Gaussian distribution, resulting in a bimodal separation of $2.22$ and a bimodality amplitude of $0.979$. These metrics strongly indicate that the output fields can be reliably categorized as either $0$ or $1$ to generate a random bitstring. The simulated system also generates one random bit with a temporal resolution on the order of \SI{15}{\femto\second}, demonstrating significant potential for its use as an all-optical, ultra-fast QRNG in the THz regime.

\section{Background and Theory}
We first analyze the OPA in the context of spontaneous parametric down-conversion (SPDC) \cite{Grice} in which the pump photons split into signal and idler photon pairs. This process obeys energy and momentum conservation:
\begin{equation}
    \omega_p = \omega_s + \omega_i,
\label{energy_conservation}
\end{equation}
\begin{equation}
    k_p = k_s + k_i,
\end{equation}
where $\omega$ and $k$ are the frequencies and wave-vectors of the photons, respectively, and the subscripts $p$, $s$, and $i$ denote the pump, signal, and idler photons, respectively. The wave-vector of a photon is characterized as
\begin{equation}
    k = \frac{2\pi n(\omega)}{\lambda},
\end{equation}
where $n(\omega)$ is the refractive index of the material at the photon's frequency $\omega$, and $\lambda$ is the photon's wavelength. Energy conservation is naturally fulfilled in the nonlinear process, but the wave-vector matching is dependent on the dispersion engineering of the nonlinear optical waveguide. Specifically, it is dependent on the phase mismatch between different propagating wave-vectors, given by
\begin{equation}
    \Delta k(\omega_s, \omega_i) = k_p (\omega_p) - k_s (\omega_s) - k_i (\omega_i),
\end{equation}
which usually has a nonzero value due to the difference in refractive indices at the pump and signal/idler frequencies. To correct for this phase mismatch, we maximize the phase-matching function $G(\Delta k)$, given by
\begin{equation}
    G(\Delta k) = \frac{1}{L} \int_{0}^{L} g(z)\exp(-iz\Delta k) dz,
    \label{eq:g(k)}
\end{equation}
where $g(z): \{0, L\} \mapsto \{-1,1\}$ is the crystal poling direction, $z$ is the spatial position along the crystal, and $L$ is the length of the crystal. The exponential term oscillates with spatial period $\frac{2\pi}{\Delta k}$, and so to maximize $G(\Delta k_c)$, where $\Delta k_c$ is the phase mismatch at our desired output wavelength, the sign of $g(z)$ is switched between $1$ and $-1$ every $\Delta z = \frac{\pi}{\Delta k_c}$, a process called “periodic poling.” This process of periodic poling ensures that the generated photons are quasi-phase-matched to the pump for efficient SPDC.

The generated biphotons can then be represented in the frequency domain as 
\begin{equation}
|\psi\rangle_{si} = \iint d\omega_s d\omega_i f(\omega_s, \omega_i) \hat{a}^{\dagger}_{\omega_s} \hat{b}^{\dagger}_{\omega_i} |0\rangle_s |0\rangle_i,
\end{equation}
where $\hat{a}^{\dagger}_{\omega_s}$ is the creation operator for the signal photon, 
$\hat{b}^{\dagger}_{\omega_i}$ is the creation operator for the idler photon, and 
$f(\omega_s, \omega_i)$ is the biphoton joint spectral amplitude with $f(\omega_s, \omega_i) \propto \alpha_{\omega_p}(\omega_s, \omega_i)G(\Delta k)$ in the low-gain regime, where $\alpha_{\omega_p}$ is the input pump spectral profile for a given pump wavelength $\omega_p$ as a function of the output signal/idler wavelengths.

\begin{figure}[!ht]
    \centering
    \includegraphics[scale=0.5]{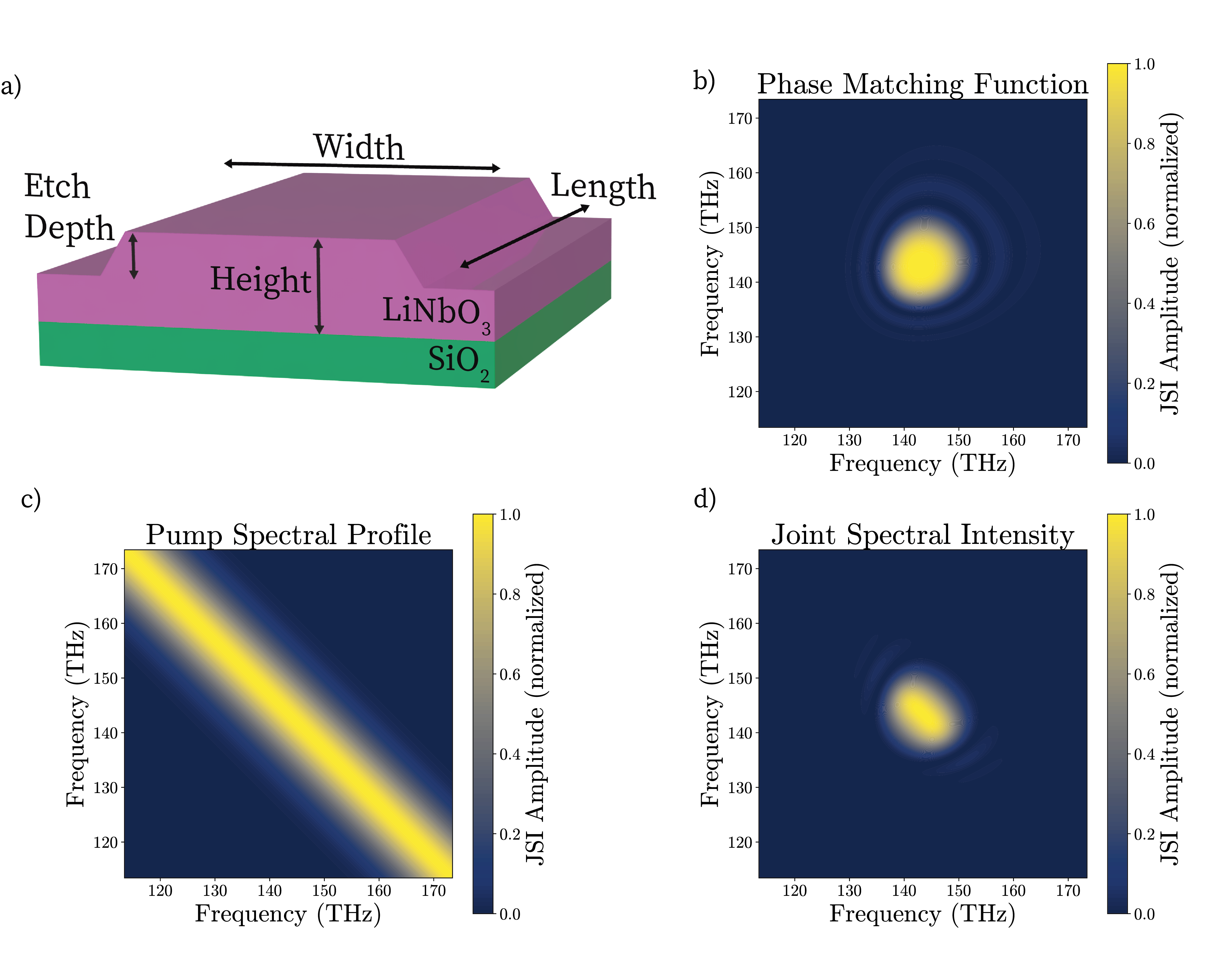}
    \caption{\textbf{a)} Diagram of waveguide dimension parameters. The primary waveguide layer is $\mathrm{LiNbO_3}$ with a buffer material of $\mathrm{SiO_2}$ underneath. \textbf{b)} Contour map of the phase-matching function for waveguide dimensions of width \SI{820}{\nano\meter}, etch depth \SI{510}{\nano\meter}, height \SI{700}{\nano\meter}, and length \SI{5}{\milli\meter}. Contour map of \textbf{c)} the pump spectral profile for a \SI{30}{\femto\second} pulse, and \textbf{d)} the product of the phase-matching function and the pump spectral profile, the joint spectral intensity (JSI), with a spectral purity of $0.61$.}
    \label{fig1}
\end{figure}

To achieve near single spectro-temporal mode operation in SPDC, and consequently in an OPA, the output state must exhibit high spectral purity. Spectral purity is maximized when only one dominant mode is present in the joint spectral intensity (JSI), $|f(w_s, w_i)|^2$. The JSI is the product of two components: the pump spectral intensity, which defines the distribution of output photon wavelength pairs based on the spectral shape of the pump and the energy conservation condition (Eq. \ref{energy_conservation}), and the phase-matching function, which describes the phase-matching distribution of the signal and idler photons as determined by the waveguide dispersion engineering. The JSI then characterizes the wavelength distribution of the photon pairs generated by the waveguide (as shown in Fig. \ref{fig1}d). The spectral purity $\mathrm{SP}$ is calculated by performing singular value decomposition on the JSI, diagonalizing it into its $n$ eigenmodes and associated eigenvalues $E_j$. Then, by normalizing the eigenvalues with $\sum_{j=1}^{n}E_j = 1$, the spectral purity is given by
\begin{equation}
    \mathrm{SP} = \sum_{j=1}^{n}E_j^2
\end{equation}

A high $\mathrm{SP}$ corresponds to a single dominant eigenmode, indicating minimal spectral correlations and near-single-mode behavior.

\section{Optimization and Simulation Methods}
\subsection*{Optimizing Waveguide Dimensions}

A waveguide is characterized by four key spatial parameters: height, etch depth, width, and length (as illustrated in Fig. \ref{fig1}a). Variations in these dimensional parameters significantly alter the waveguide's dispersion profile, influencing the propagation characteristics of the pump pulse and the OPA process \cite{ledezma2022}. Specifically, the geometric confinement of the waveguide structure substantially affects the effective refractive index of the waveguide modes. Consequently, adjustments to the waveguide dimensions lead to modifications in the phase-matching conditions, directly altering the JSI.

Using the Nelder-Mead optimization algorithm \cite{neldermead}, we optimize for high spectral purity under this parameter space. For each waveguide dimension, the refractive indices for a range of wavelengths are calculated using an analytical model for the dispersion profile \cite{Ledezma}, yielding the corresponding phase-matching function. With this information, we can then compute the spectral purity. To effectively search the high-dimensional parameter space, we employ a sweeping strategy across a range of realistic initial conditions for the waveguide dimensions, using a pump pulse with a wavelength of \SI{1045}{\nano\meter} based on a previously physically realized design \cite{ledezma2022}. For each initial condition, the optimization algorithm is executed to identify a corresponding local maximum in the vicinity. By selecting the largest of the local maxima obtained from these sweeps, we approximate the global maximum for the spectral purity. The accuracy of this approximation is directly influenced by the density of the initial conditions sampled and the extent of the parameter space explored. A finer set of initial conditions improves the likelihood of converging towards the true global maximum.

\subsection*{Optimizing the Poling Design
}
\begin{figure}[!ht]
    \centering
    \includegraphics[scale=0.69]{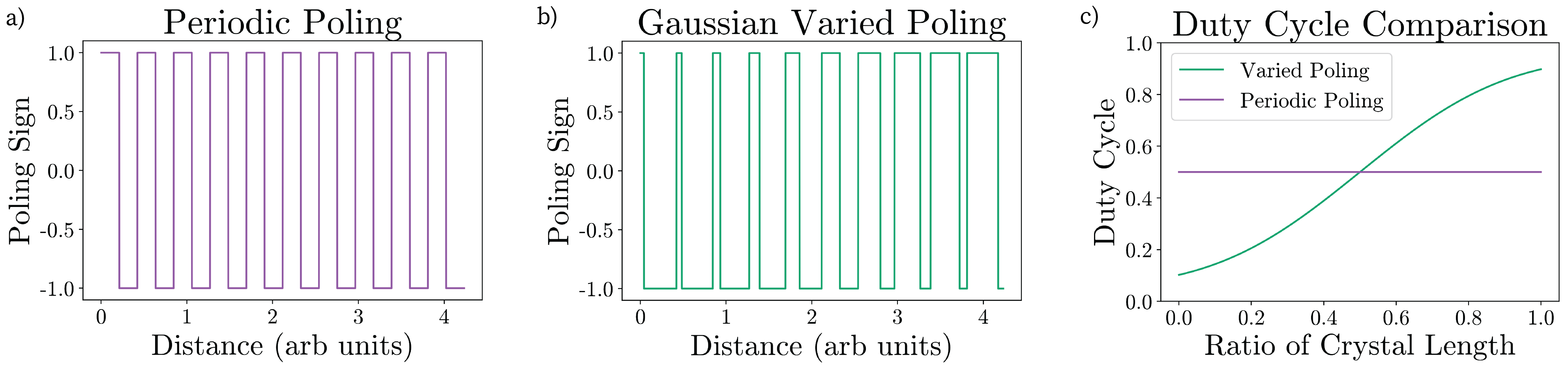}
    \caption{\textbf{a)} Periodic poling with a constant duty cycle of $0.5$, \textbf{b)} varied poling according to a Gaussian error function duty cycle with parameters $a = 0.5, b = 0.9$, $c=0.45$ and \textbf{c)} duty cycle comparison between the two poling designs.}
    \label{fig2}
\end{figure}

The periodic poling design, motivated by Eq. \ref{eq:g(k)}, yields a phase-matching function characterized by a $\mathrm{sinc}$ function profile in the frequency domain, which exhibits side lobes that contribute to reduced spectral purity (as seen in Figs. \ref{fig1}b, c). To mitigate this, the poling design can be modified to yield a phase-matching function approximating a Gaussian profile in the frequency domain by varying the duty cycle of the poling pattern between $+1$ and $-1$. The duty cycle refers to the fraction of the poling period during which the crystal orientation aligns with $+1$. As first demonstrated in \cite{Dixon}, a near-Gaussian function in momentum space can be achieved by varying the duty cycle, $d(z)$ according to a Gaussian error function along the length of the crystal (illustrated in Fig. \ref{fig2}b), such that
\begin{equation}
    d(z) = a \times (1+b \times \erf(\frac{z-L/2}{c})),
\end{equation}
where $z$ is the distance along the crystal, $\erf(x)$ is the Gaussian error function defined as $\erf(x) = \frac{2}{\sqrt{\pi}}\int_0^x e^{-t^2} dt,$
$L$ is the length of the crystal, and $a$, $b$, and $c$ are optimizable variables that vary the shape of the Gaussian duty cycle. For our purposes, $c$ can be fixed to an arbitrary value and we only need to optimize the values of $a$ and $b$ (an example of a Gaussian error function is shown through the green plot in Fig. \ref{fig2}c). 

This process suppresses the side lobes present in the phase-matching function (as shown in Fig. \ref{fig3}d, where the side lobes present in Figs. \ref{fig1}b, c are not observed). By applying the Nelder-Mead optimization algorithm with the added parameters $a$ and $b$ ($c$ is arbitrarily fixed) that vary the Gaussian error function, the highest spectral purity achievable with a $15$-fs pump pulse increased from $0.723$ to $0.937$.

\subsection*{Implementing the Complete Group Velocity Matching Condition}
Based on a recent study optimizing spectral purity on a type-1 SPDC platform \cite{horoshko2024}, we implement a complete group velocity matching (cGVM) condition by sweeping wavelength values to find the wavelength at which the pump and signal/idler pulses traveling in the crystal have the same group index, and hence the same group velocity. This maximizes spectral overlap between the pulses, suppressing temporal walk-off and increasing OPA generation efficiency. We find that an iterative process between performing optimization of waveguide dimensions and applying cGVM matching yields increased spectral purity values. This is partly due to the centering of the phase-matching function along the pump spectral profile so that the JSI maintains the shape of the phase-matching function. This is seen in Figs. \ref{fig3}b, c where the former phase-matching function is uncentered and the latter is centered with respect to the pump spectral profile. Modifying the pump wavelength $(\lambda /2)$ from $1045$ nm to the cGVM matching $1115$ nm (Fig. \ref{fig3}a), and subsequently repeating the optimization increased the spectral purity from $0.937$ to $0.982$.

\section{Results and Applications}
\label{results}
\subsection*{Optimized Spectral Purity Values}
\begin{figure}[!ht]
    \centering
    \includegraphics[scale=0.7]{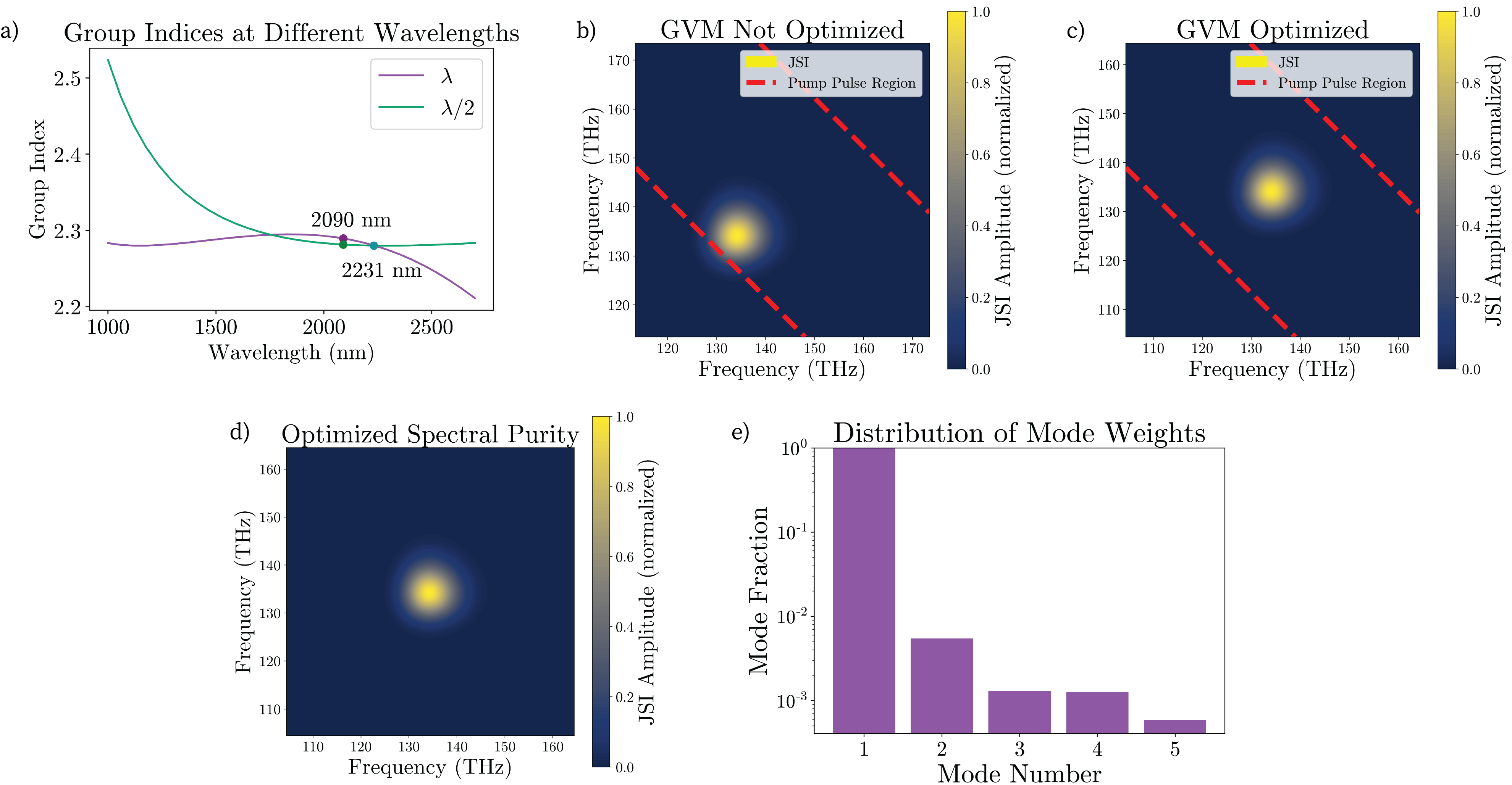}
    \caption{\textbf{a)} Plot of group velocities over wavelengths $\lambda$ and $\lambda/2$ to find the intersection for cGVM matching. Our initial signal wavelength was at $2090$ nm which we then shifted to the intersection point at $2231$ nm. \textbf{b)} JSI of a waveguide optimized for dimensions and poling design but not cGVM matched (spectral purity $0.937$), and \textbf{c)} JSI of the same waveguide with cGVM matching (spectral purity $0.940$). In this case, the pump pulse frequency is centered on the JSI. {\textbf{d)} Joint spectral intensity of the optimized waveguide with dimensions of width \SI{983}{\nano\meter}, etch depth \SI{502}{\nano\meter}, height \SI{700}{\nano\meter}, and length \SI{6}{\milli\meter} for a pulse of \SI{10}{\femto\second} (spectral purity $0.982$). \textbf{e)} Distribution of mode weights of the 5 most dominant modes present in the JSI in log scale, with a dominant mode fraction of $0.991$.}}
    \label{fig3}
\end{figure}

We optimize the waveguide dimensions, the poling design, the input pump wavelength, and the pump pulse duration to obtain a spectral purity value of $0.982$, the highest reported simulated value for a type-0 OPA. After calculating the JSI of the output pulse, we plot the weight of each mode present in the SVD of the JSI matrix. The mode distribution, displayed in Fig. \ref{fig3}e, indicates that the dominant mode has a mode fraction of $0.991$.

When testing this waveguide design in our semiclassical simulation, we were limited by the fact that the design required a $10$-fs pump pulse to achieve this purity. At such a short pulse duration, the pump spectrum in the frequency domain overlaps with the signal and idler frequencies, effectively seeding the OPG process. This compromises the desired condition of amplifying only vacuum fluctuations. To address this, we optimized the waveguide design for a slightly longer pump pulse duration of \SI{15}{\femto\second}. When constrained by design parameters for which our simulation could compute refractive indices across all required wavelengths, we achieved a maximum spectral purity of $0.978$. It is important to note that this optimized design does not fully adhere to physical constraints, such as a minimum duty cycle or a minimum input pump pulse duration. However, it serves as a guideline for achievable spectral purity for type-0 optical parametric amplification implemented in TFLN waveguides. This high spectral purity demonstrates the potential of this scheme to generate single spectro-temporal mode biphotons and squeezed vacuum states for various quantum information applications.

\subsection*{Application as a Quantum Random Number Generator}

Unlike multimode optical parametric generation \cite{ledezma2022, jankowski2022} where the pulse-to-pulse spectrum can vary substantially, a property of single-mode OPG is that the identical pulse-to-pulse spectrum at degeneracy projects either onto a $0$ or $\pi$ phase when measured. As such, by associating these phases with $0$ and $1$ outputs, respectively, we can utilize the output signal for a binary QRNG.

\begin{figure}[!ht]
    \centering
    \includegraphics[scale=0.75]{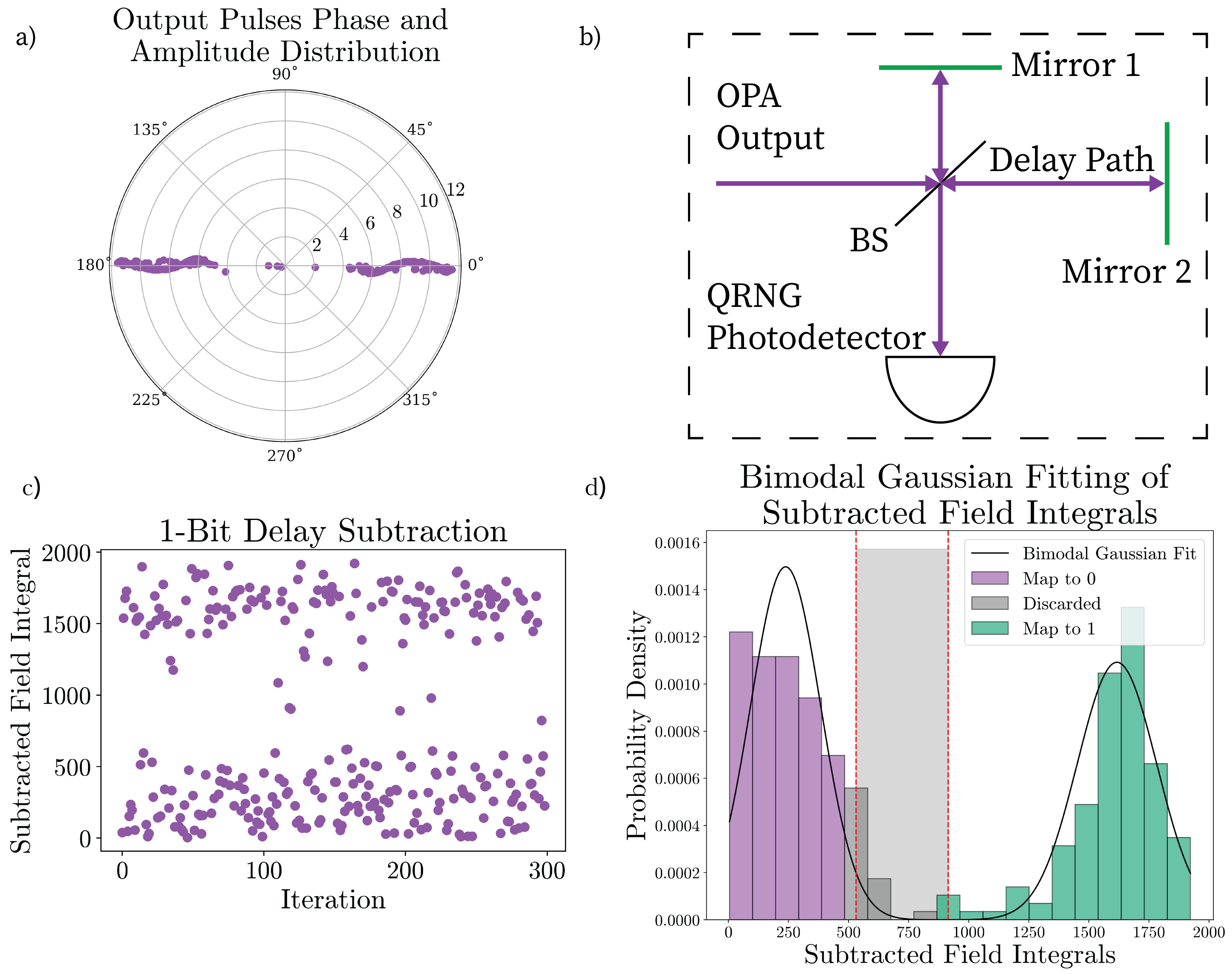}
    \caption{\textbf{a)} Polar plot of the phase and amplitude of 300 simulated output pulses at an input pump pulse power of \SI{29}{\pico\joule} showing bimodality in phase distribution and relatively stable amplitudes. \textbf{b)} Proposed experimental setup for applying the OPA waveguide to a QRNG, where fields produced through the OPA process interfere with subsequent incoming fields within a delay line interferometer. The OPA output goes through a beam splitter, and the round trip length difference of the interferometer arms is equal to the separation of the OPA outputs. The resultant constructive/destructive signal measured by the QRNG photodetector would correspond to a digital bit. \textbf{c)} Scatter distribution of the one-bit delay subtracted field integrals of 300 simulated output pulses, and \textbf{d)} histogram distribution of the subtracted field integrals fitted to a bimodal Gaussian curve with a bimodal separation of $2.22$ and a bimodality amplitude of $0.979$. $51.84\%$ of the outputs are mapped to binary $0$, another $42.81\%$ of the outputs are mapped to binary $1$, and $5.35\%$ of the outputs are discarded.}
    \label{fig4}
\end{figure}

To characterize the application of our waveguide design as a QRNG, we utilize a previously developed nonlinear optics package \cite{Ledezma} that takes an input pump pulse and models its progression through the waveguide based on a Fourier split-step solution of the nonlinear envelope equation in the frequency domain (details in \cite{ledezma2022}). In doing this, we take into account linear dispersion effects and nonlinear effects to compute the output field. We also inject a semiclassical model of quantum noise into the solver to emulate the vacuum fluctuations. The output can then be analyzed in both the time domain and frequency domain (simulation visualized in Fig. \ref{supfig1}), and we can hence extract the output field's phase and amplitude, as shown in Fig. \ref{fig4}a for $300$ outputs.

To evaluate the system's viability as an ultra-fast QRNG, we simulate a one-bit delay protocol to measure the relative phases of consequent pulses. In this protocol, consecutive optical fields are subtracted from each other, leading to either constructive or destructive interference. As illustrated in Fig. \ref{fig4}b, the experimental setup would involve interfering incoming fields with subsequent fields produced by the high-gain depleted single-mode OPA process using a delay line interferometer. The intensity of the pulses at the output of the interferometer, referred to as subtracted field integrals in our simulation, would be either high or low depending on whether the consequent pulses are in-phase or out-of-phase. This output would then correspond to a register of $0$ or $1$ based on its value relative to a cutoff threshold. The cutoff threshold is defined as the midpoint of the largest difference between any two subtracted field integrals. Following the method of digital signal generation, we implement a buffer zone with a range spanning $20\%$ of the subtracted field integral axis and centered on the cutoff threshold (shaded domain area in Fig. \ref{fig4}d). Subtracted field integrals within this range are discarded to maintain signal accuracy. By repeating this process, a concatenated bitstring is generated, representing a random number. 

Tuning the energy of the input pump pulse yielded different distributions of output phase and amplitudes (see Fig. \ref{supfig2} for some examples). As such, we optimized for high resolvability of binary outputs by employing a bimodal Gaussian fitting approach to the subtracted field integrals. Specifically, we targeted high bimodal separation and bimodality amplitude (see Sec. \ref{sup_bimodal} for details about the fitting methodology). We swept over varying input energy values to optimize for high bimodality (displayed in Fig. \ref{supfig3}), and our results indicate that an input pump pulse energy of \SI{29}{\pico\joule} provides optimal performance. With $300$ pulse runs, we obtain a binary distribution of subtracted field integrals that, when fitted to a bimodal Gaussian distribution with a buffer zone, has a bimodal separation of $2.22$ and a bimodality amplitude of $0.979$. This indicates strong support for the bimodality and hence the resolvability of the data set to binary values. Furthermore, $51.84\%$ of outputs resolved to binary $0$, $42.81\%$ of outputs resolved to binary $1$, and $5.35\%$ were discarded. While this output does not demonstrate a $1:1$ ratio of generation of $0$ and $1$ bits, one could implement a randomness extractor protocol such as the Von Neumann extractor \cite{vonNeumann} to transform the biased output into a distribution with a uniform ratio. Furthermore, when sweeping at higher energy values, we obtain a near $1:1$ generation ratio with acceptable bimodality attributes at $\SI{46.4}{\pico\joule}$ (see Sec. \ref{1_1_generation}). This candidate provides an alternative pathway for unbiased number generation without requiring additional randomness extraction protocols.

These results highlight the potential of the proposed waveguide as a QRNG. With simulated input pulses of \SI{15}{\femto\second} duration, the waveguide is capable of functioning as an all-optical, ultra-fast QRNG in the THz regime.

\section{Conclusion}
In this work, we successfully optimized the design of a thin-film lithium niobate waveguide to achieve a high simulated spectral purity value in a type-0 degenerate optical parametric amplifier (OPA). This design overcomes the challenge of multimode operation of ultrashort pulse OPAs for quantum information systems. By varying the waveguide dimensions, the poling design, the pump pulse duration, and the input pump wavelength to achieve complete group velocity matching, we achieved a theoretical spectral purity of 0.982. We also acknowledge that our optimization regime has been fairly narrow, as we have only varied pump pulses, waveguide geometry, and poling duty-cycle under fixed poling period conditions. It is likely that more comprehensive optimization strategies can lead to even better performances.

Beyond enhancing spectral purity, we demonstrated a practical application of our optimized waveguide as a quantum random number generator (QRNG). Using nonlinear time-domain simulations, we observed binary probabilistic phase behavior in the macroscopic output of the OPA, which we mapped to binary states to simulate an ultra-fast, all-optical QRNG capable of generating random bits with a temporal resolution of approximately \SI{15}{\femto\second}. These results show a path for the development of ultrafast quantum circuits using single-mode nanophotonic OPAs.

\section*{Acknowledgement}
The authors gratefully acknowledge support from ARO grant no. W911NF-23-1-0048, NSF grant no. 1846273, 1918549, 2139433, AFOSR award FA9550-23-1-0755, the center for sensing to intelligence at Caltech, the Alfred P. Sloan Foundation, and NASA/JPL.

\printbibliography
\clearpage

\appendix
\renewcommand{\thefigure}{S\arabic{figure}}
\renewcommand{\thesection}{S\arabic{section}}
\setcounter{figure}{0} 
\setcounter{section}{0}
\section*{\huge Supplementary}
\section{Pulse Propagation Simulation}
\begin{figure}[!ht]
    \centering
    \includegraphics[scale=0.9]{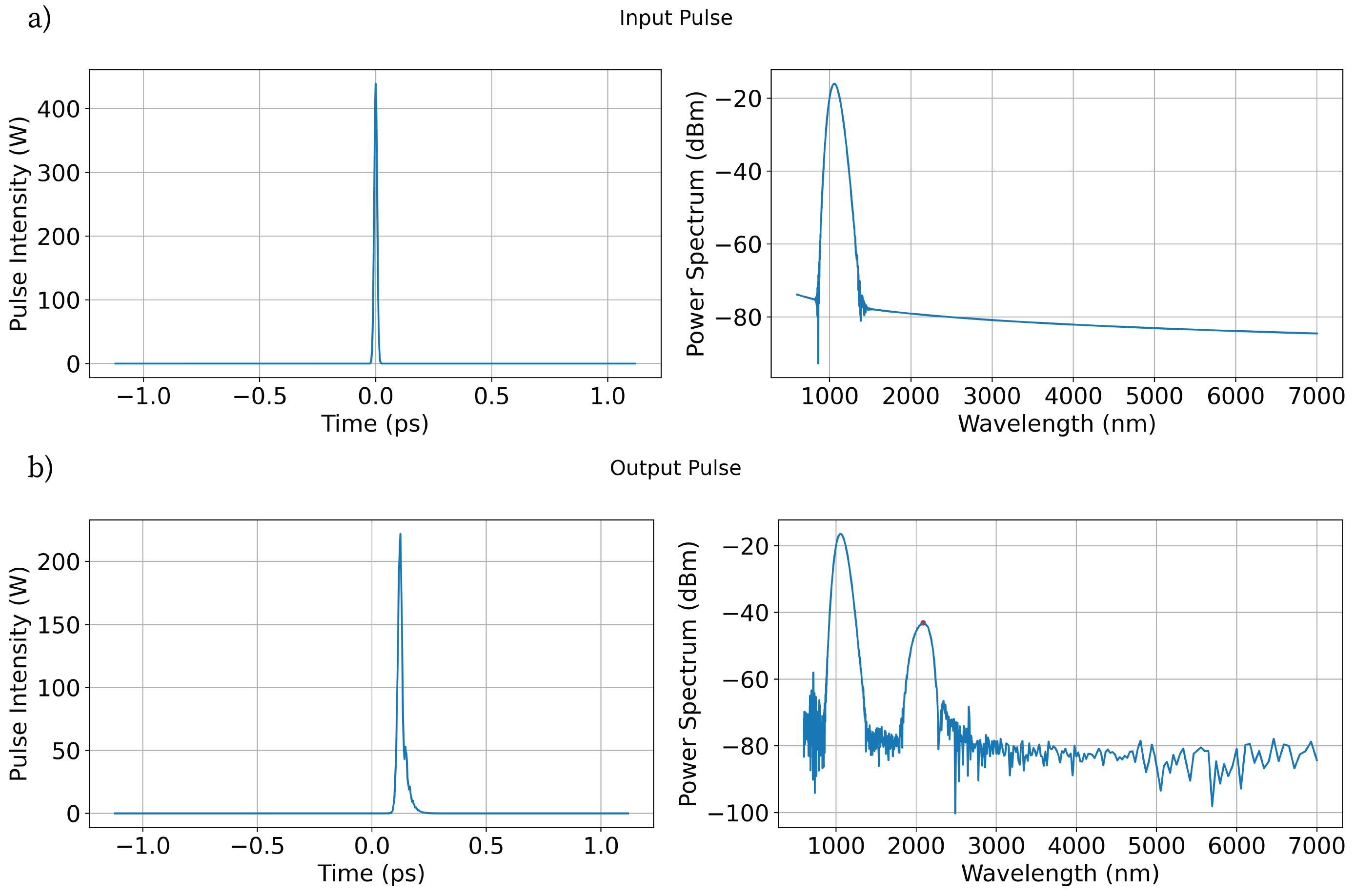}
    \caption{\textbf{a)} Input pump pulse in the time domain and wavelength domain peaked at $\SI{1045}{\nano\meter}$, \textbf{b)} output pump pulse in the time domain and wavelength domain with an additional peak at $\SI{2090}{\nano\meter}$ due to the optical parametric amplification process.}
    \label{supfig1}
\end{figure}

To simulate the propagation of a pulse through our simulated waveguide, we employed a previously developed nonlinear time-domain simulation package \cite{Ledezma}, which tracks the propagation of an input pump pulse through the waveguide, modeled in small spatial steps. An example of a $7$ pJ input pulse and its corresponding output is shown in Fig. \ref{supfig1}.

The left column of the figure presents the input and output pulses in the time domain, providing a temporal view of the signal evolution. More notably, the right column depicts the corresponding pulses in the wavelength domain. The spectrum of our input pulse is centered at $\SI{1045}{\nano\meter}$. In contrast, the output pulse displays an additional peak centered at $\SI{2090}{\nano\meter}$ due to the optical parametric amplification (OPA) process. Additionally, the output spectrum exhibits increased noise artifacts across all wavelengths. This is due to the addition of a semiclassical model of quantum zero-field noise of $\frac{1}{2}\hbar\omega$ inserted at each Fourier split-step during the simulation. We filter out the OPA components of the output pulse in the time domain and use them as the output fields for our proposed QRNG scheme.

\clearpage

\section{Bimodal Gaussian Fitting and Measures of Resolvability}
\label{sup_bimodal}
To characterize whether our interfered fields could be resolved to binary $0$ or $1$ in Sec. \ref{results}, we fitted the subtracted field integral data to a bimodal Gaussian distribution. This is a type of probability distribution that consists of two distinct Gaussian distributions combined into one model. Each of the two distributions represents a mode in the overall distribution. 

The probability density function of a bimodal Gaussian distribution is typically represented as a weighted sum of two individual Gaussian distributions. Mathematically, it can be written as:

\[
f(x) = w_1 \cdot \frac{1}{\sqrt{2\pi\sigma_1^2}} \exp\left( -\frac{(x - \mu_1)^2}{2\sigma_1^2} \right)
+ w_2 \cdot \frac{1}{\sqrt{2\pi\sigma_2^2}} \exp\left( -\frac{(x - \mu_2)^2}{2\sigma_2^2} \right)
\]

where \( \mu_1 \) and \( \mu_2 \) are the means of the two Gaussian components, \( \sigma_1 \) and \( \sigma_2 \) are the standard deviations of the two Gaussian components, \( w_1 \) and \( w_2 \) are the weights of each Gaussian component, respectively, which satisfy \( w_1 + w_2 = 1 \), and \( x \) is the random variable. For the fitting in Sec. \ref{results}, we obtain values $\mu_1 = 238.24$, $\sigma_1 = 145.98$, and $\mu_2 = 1616.79$, $\sigma_2 = 165.19$.

We use two measures of bimodal resolvability, the bimodal separation and the bimodality amplitude. The bimodal separation is a measure of how distinct or separated the two modes of a bimodal Gaussian distribution are. It is calculated using the means and standard deviations of the two Gaussian components. The equation for bimodal separation \( S \) is:
\[
S = \frac{|\mu_1 - \mu_2|}{2(\sigma_1 + \sigma_2)}
\]
A higher value of \( S \) implies that the two peaks are well-separated and easily distinguishable, whereas a lower value of \( S \) suggests that the peaks are closer together and may overlap, making the distribution less bimodal. Generally, a value larger than $2$ is considered to be supportive of distinguishability, and so the simulated value of $2.22$ for the bimodal separation provides strong support for our claim of resolvability.

Next, the bimodality amplitude \( A_B \) is a measure of the depth of the valley between the two modes relative to the height of the smaller peak. It compares the amplitude of the smaller "mode" peak with the amplitude of the "antimode" (the lowest point between the two peaks). The equation for bimodality amplitude is:

\[
A_B = \frac{A_1 - A_{\text{AM}}}{A_1}
\]
where $A_1$ is the value of the smaller mode peak and $A_{\text{AM}}$ is the amplitude of the antimode. A value of \( A_B \) between $0.7$ and $1$ indicates a well-defined bimodal distribution with a clear valley between the peaks (and hence better resolvability), and so the simulated value of $0.979$ also strongly supports our claim of resolvability.

\clearpage

\section{Pulse Output Profile Over Various Pump Energies}
\label{energy_sweep}

\begin{figure}[!ht]
    \centering
    \includegraphics[scale=0.9]{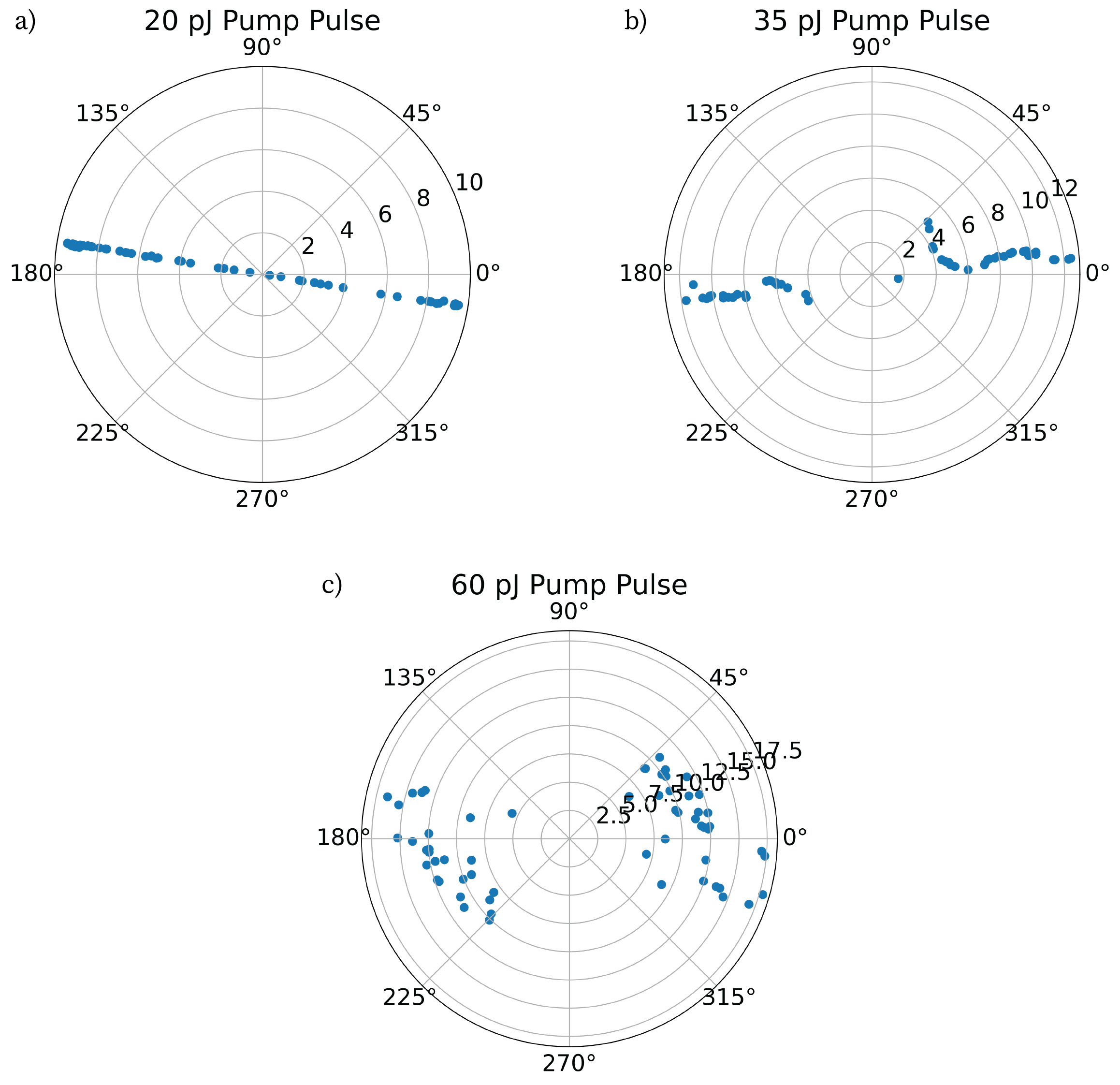}
    \caption{60 output pulse distributions at \textbf{a)} $20$ pJ showing very high phase coherence but also high amplitude variance, \textbf{b)} $35$ pJ showing decreasing phase coherence and also decreasing amplitude variance, \textbf{c)} $60$ pJ showing even lower amplitude variance but the phase coherence is too low for it to be viable for one-bit delay subtraction.}
    \label{supfig2}
\end{figure}

As stated in Sec. \ref{results}, different pulse energies yielded different distributions of output phase and amplitudes. Specifically, energy values lower than our optimal value exhibited a strong phase coherence where points were either in phase or $\pi$ out of phase with each other but also exhibited very high amplitude variance (as seen in Fig. \ref{supfig2}a). This amplitude variance excludes the usage of this regime for a 1-bit delay subtraction scheme. On the other hand, at higher energy values the amplitude variance starts to decrease, but the phase variance starts to increase (Fig. \ref{supfig2}c), also excluding the usage of this regime for a 1-bit delay subtraction scheme. As such, we observe a competing effect of decreasing phase coherence and decreasing amplitude variance when increasing pump energy. Hence, we needed to sweep the energy values to find the optimal energy for the bimodality of the 1-bit delay subtraction, which we found to be at $29$ pJ. 

\clearpage

To do a systematic sweep over energy values to find the optimal value, we searched for high bimodal separation and bimodality amplitude of the bimodal Gaussian fitting of the data. It is important to emphasize that the values in this sweep are generated without incorporating discarded buffer zones in order to avoid inaccurate comparisons, resulting in values that are lower than the optimal level reported in the main text. We observe a general trend of increasing values of bimodality parameters and a subsequent dip after the chosen value of $\SI{29}{pJ}$ near the maximum point for both.

\begin{figure}[!ht]
    \centering
    \includegraphics[scale=0.95]{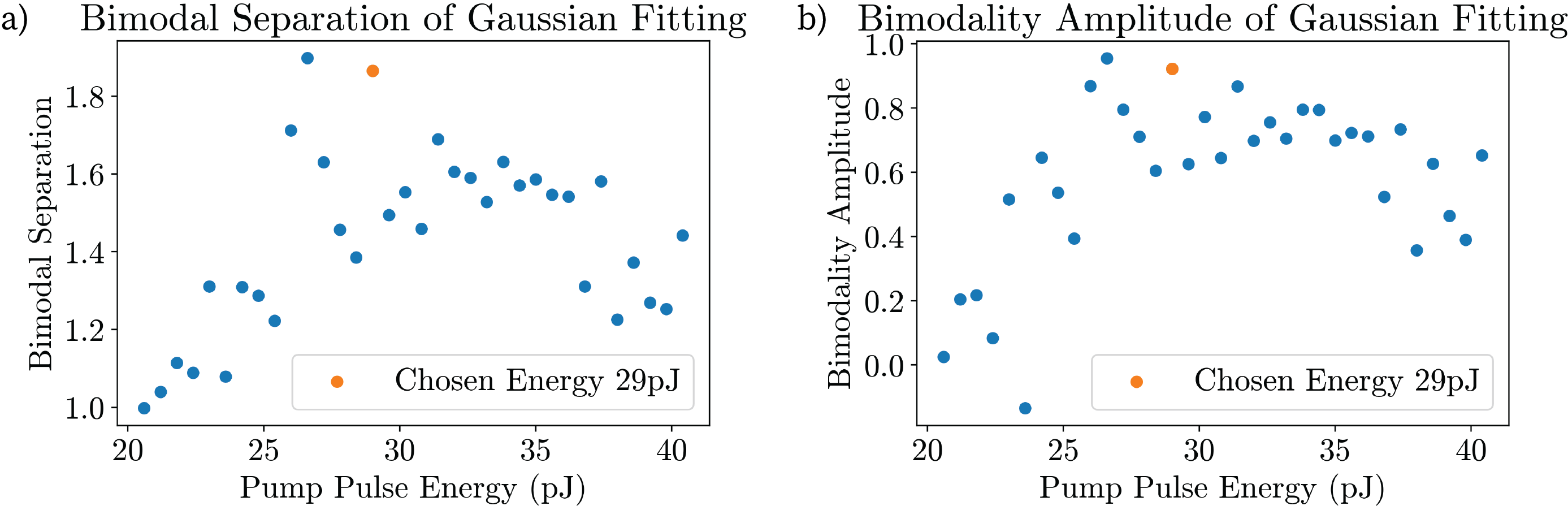}
    \caption{Sweeping over pump pulse energies to evaluate the \textbf{a)} bimodal separation and \textbf{b)} bimodality amplitude of the bimodal Gaussian fits. The chosen energy of $\SI{29}{pJ}$ is in orange.}
    \label{supfig3}
\end{figure}

\clearpage

\section{Energy Candidate for 1:1 Bit Generation Ratio}
\label{1_1_generation}

\begin{figure}[!ht]
    \centering
    \includegraphics[scale=1]{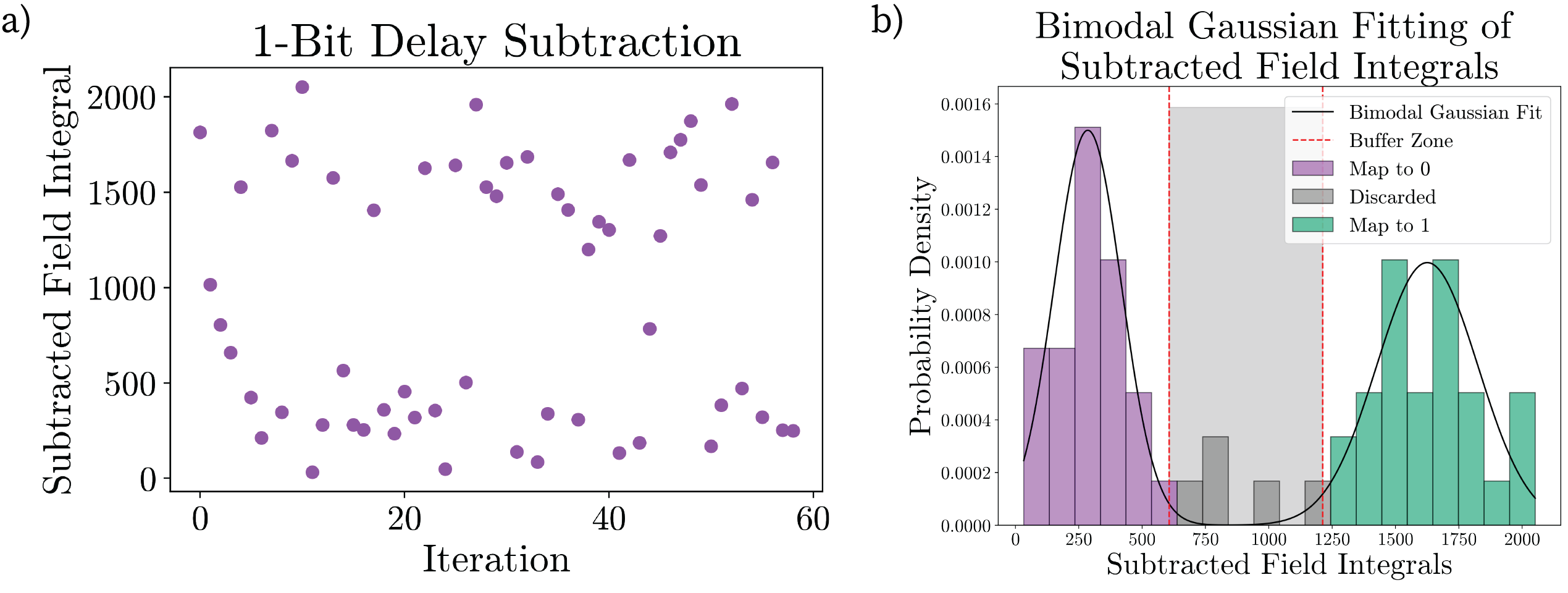}
    \caption{QRNG data for an input energy of $\SI{46.4}{\pico\joule}$. \textbf{a)} Scatter distribution of the one-bit delay subtracted field integrals of 60 simulated output pulses, and \textbf{b)} histogram distribution of the subtracted field integrals fitted to a bimodal Gaussian curve with a bimodal separation of $2.01$ and a bimodality amplitude of $0.879$. $45.76\%$ of the outputs are mapped to binary $0$, another $45.76\%$ of the outputs are mapped to binary $1$, and $8.47\%$ of the outputs are discarded.}
    \label{supfig4}
\end{figure}

As indicated in Sec. \ref{results}, we swept for bimodality at higher energy values and found a candidate at $\SI{46.4}{\pico\joule}$ with good bimodality parameters that outputs $0$ and $1$ bits at a $1:1$ generation ratio. With $60$ pulse runs, we obtain a binary distribution of subtracted field integrals with a bimodal separation of $2.01$ and a bimodality amplitude of $0.879$ (the data is displayed in Fig. \ref{supfig4}). This indicates strong support for bimodality, although not as pronounced as the $\SI{29}{\pico\joule}$ data presented in the main text. However, $45.76\%$ of outputs resolved to $0$, another $45.76\%$ of outputs resolved to $1$, and $8.47\%$ of outputs were discarded. Hence, when considering only the field integrals mapped to binary values, this energy level yields a uniform $1:1$ generation ratio of $0$ and $1$ bits.

\end{document}